\documentclass[aps,prl,twocolumn,superscriptaddress,proof,showpacs]{revtex4}

\usepackage{amsmath,amssymb}
\usepackage{graphicx}
\usepackage{dcolumn}

\newcommand{\br}{\mathbf r}

\newcommand{\bp}{\mathbf p}

\begin{document}


\title{Path integral study of the role of correlation in exchange coupling \\
 of spins in double  quantum dots and optical lattices}


\author{Lei Zhang}
\affiliation{Deptartment of Physics, Arizona State University, Tempe, AZ 85287-1504, USA}
\author{M. J. Gilbert}
\affiliation{Microelectronics Research Center, University of Texas,  Austin, Texas 78758, USA}
\affiliation{Department of Electrical and Computer Engineering,
University of Illinois, 
Urbana, IL 61801,  USA}
\author{Jesper Goor Pedersen}
\affiliation{
DTU Fotonik, Department of Photonics Engineering, Technical University of Denmark, Kongens Lyngby, Denmark }
\author{J. Shumway}
\email[]{shumway@mailaps.org}
\homepage[]{http://phy.asu.edu/shumway}
\affiliation{Deptartment of Physics, Arizona State University, Tempe, AZ 85287-1504, USA}

\date{\today}

\begin{abstract}
We explore exchange coupling of a pair of spins in a double dot and in an optical lattice.
Our algorithm uses the frequency of exchanges in a bosonic
path integral, evaluated with Monte Carlo. This algorithm is simple enough to 
be a ``black box'' calculator, yet gives insights into the role
of correlation through two-particle probability densities, visualization of instantons, and
pair correlation functions. We map the problem to Hubbard model and see that 
exchange and correlation  renormalize
the effective parameters, dramatically lowering $U_r$ at larger separations.
\end{abstract}


\maketitle

Lattice models are popular in solid state physics,
often serving as simple models for atomic orbitals,
 especially in the theory of magnetism \cite{Mattis:1981}.
Quantum dot (QD) arrays
and optical lattices (OL) have emerged as new realizations
of lattices. These artificial lattices are candidates for 
quantum computers, where spins on exchange-coupled QDs 
act as qubits for universal quantum computation~\cite{Loss:1998,Burkard:1999}.
A basic concept of lattice models is the intersite exchange,
in which virtual hopping leads to an effective spin coupling of neighboring sites.
A two-site model is one of the simplest quantum problems, yet the quantitative mapping
from a three-dimensional model of a recent double QD or OL experiment
to an effective two-site Hamiltonian has many
subtleties requiring careful treatment of exchange and 
correlation \cite{Mattis:1981,Burkard:1999,Pedersen:2007}.

In this Letter we develop a simple path integral Monte Carlo (PIMC) 
approach for extracting accurate singlet-triplet splitting from a two- or three-dimensional 
continuous model.
This two-particle problem has been previously solved with direct diagonalization (DD)
methods with a careful choice of basis functions  \cite{Pedersen:2007} and is amenable to 
variational or diffusion quantum Monte Carlo (QMC) \cite{Ghosal:2006}.
However, the simple and elegant  PIMC approach is a
more direct solution without variational bias or basis-set issues and offers 
theoretical insights into this important problem. 
We first show that the splitting energy, $J$, is easily extracted from the average 
permutation of the two-particle path integral, even when $J\ll k_BT$. 
This PIMC algorithm can be a black-box 
calculator, providing accurate numerical estimates of $J$ for 
technologically relevant models of quantum dots or optical lattices, with arbitrary interactions
and confinement potentials.  The path integral also
allows us to ask questions about the correlated behaviour
of the electrons. For example, do the electrons exchange
across the barrier simultaneously, or do they briefly double
occupy the dot?  Or, how does the motion of
one electron over the barrier correlate with the location of
the other electron? We answer these questions by viewing
representative trajectories (instantons) in the path integral, and, more
quantitatively, by calculating pair correlation functions. 
Next, we include the effects of a magnetic field on electron exchange in double QDs
using a Berry's phase for the magnetic flux enclosed by the electron paths \cite{berry:1984}.
Finally we use the method to model recent experiments of exchange coupled atoms
in an optical trap, demonstrating its broad utility \cite{Trotzky:2008}.

The mapping from a continuous model with interacting
particles to a lattice model introduces subtle complications.
For a non-interacting system it is reasonable to reduce the 
Hilbert space to include just one orbital per site, such as
a Wannier function centered on each potential minimum.
The non-interacting many-body ground state 
is a product state of these single particle orbitals, and
low-lying excited states are also spanned by this basis, so
an effective lattice model is an excellent approximation.
Interactions are typically added to this lattice model, often
as on-site energies, $U$, or intersite terms,  $V$. For small $t$,
this gives the well-known $J=-4t^2/(U-V)$.

There can be a serious flaw when considering interactions
in this order, by first mapping to a lattice then adding interactions.
When interactions are added to the continuum Hamiltonian,
correlation enters as virtual excitations to higher
energy orbitals. At first this seems insignificant,
since there may be still a one-to-one mapping to an effective lattice model.
But, when choosing effective lattice parameters, one must remember
that many-body states in the continuum model have quantum fluctuations
that are simply not present in the lattice model.

As a specific example, consider two electrons in two coupled
QDs. This system is often represented at a two-site
Hubbard model, where the sites represent the $1s$ ground states
of the quantum dots. Correlation terms involve virtual excitation
of the electron to the $2p_x$. and $2p_y$ states of the dots.
These quantum fluctuations give rise to van der Waals attraction, in 
addition to the usual mean-field Coulomb repulsion. Van der
Waals attraction and other correlations renormalize
the interaction parameters to new values, $U_r$ and $V_r$.

When we consider hopping between sites, more 
complications emerge. The hopping barrier has contributions from both the
external potential and electron-electron interactions.
While the mean-field Hartree contribution can simply be
added to the effective potential, the fluctuating part
is not so trivial. In the transition state,
an electron passes over a barrier whose height has
quantum fluctuations. Thus we expect interactions to renormalize
the hopping constant, $t_r$. At the Hartree-Fock level, 
Hund-Mulliken theory already predicts a renormalized 
$t_r$ and $U_r$ due to long-range exchange. However,
the neglect of correlation in Hund-Mulliken theory can lead to catastrophic 
failure at intermediate dot separations \cite{Pedersen:2007}.
Our PIMC approach includes all correlation and illuminates its role
in barrier hopping with the concept of instantons.

We start with the two-dimensional model for the GaAs double quantum dot
studied in Ref.~[\onlinecite{Pedersen:2007}],
\begin{equation}
H=\frac{\bp_1^2}{2m^*}+\frac{\bp_2^2}{2m^*}
+\frac{e^2}{\epsilon|\br_1-\br_2|}
+V_{\text{ext}}(\br_1) +V_{\text{ext}}(\br_2),
\end{equation}
with $m^*=0.067 m_e$ and $\epsilon=12.9$. 
The external potential comes from two piecewise-connected
parabolic potentials,
\begin{equation}
V_{\text{ext}}(\br)=\textstyle\frac{1}{2}m\omega_0\{ \min [(x-d)^2,(x+d)^2] +y^2\},
\end{equation}
with minima at $x=\pm d$. We report $d$ relative to the oscillator length $r_0=\sqrt{\hbar/m\omega_0}$.
The two lowest energy two-electron states are spatially symmetric and
anti-symmetric under exchange, with energies $\varepsilon_+$ and $\varepsilon_-$,
respectively. The exchange coupling, $J=\varepsilon_-
-\varepsilon_+$, has been calculated previously using DD on a basis of 
Fock states built from seven single particles states~\cite{Pedersen:2007}.
Much care was taken to test convergence with the number of states and careful
evaluation of coulomb matrix elements. We note that the same quality of DD calculation in 
three dimensions would typically take more single particle  states.

QMC techniques give
essentially exact answers to many problems
without basis set convergence issues, and often work just as
easily in two or three dimensions. PIMC is nice for QD
problems~\cite{Harowitz:2005b} because it does not require a trial wavefunction.
However, a direct calculation of either $\varepsilon_+$ or $\varepsilon_-$ with PIMC 
often has a large statistical errors in energy ($\sim 1$~meV in QDs). 
Instead, we use particle exchange statistics to estimate energy difference 
$J$ to high accuracy ($\sim 1$~$\mu$eV) in PIMC.

To calculate J, we split the partition function
into two parts that are either spatially symmetric and antisymmetric under exchange, $Z=Z_+ + Z_-$.
These two terms can be expressed as a symmetrized and antisymmetrized 
imaginary-time
path integral,~\cite{Feynman:1972,Ceperley:1995}
\begin{equation}
Z_\pm = \frac{1}{2!}\sum_{P=\mathcal{I},\mathcal{P}}
(\pm1)^P\int\mathcal{D}R(\tau)e^{-\frac{1}{\hbar}S_E[R(\tau)]}.
\end{equation}
This is a sum over all two-particle worldlines $R(\tau)$
with the boundary condition $R(\beta\hbar)=PR(0)$, where the operator $\mathcal{P}$ swaps
the particles. The symbol $(\pm1)^P$
takes on the values $(\pm1)^\mathcal{I}=1$, $(+1)^\mathcal{P}=1$,
and  $(-1)^\mathcal{P}=-1$.
At low temperature, only the two lowest states contribute to the partition functions, so 
$Z_{\pm}=e^{-\beta\varepsilon_\pm}$. Then,
\begin{equation}
e^{-\beta J}=\frac{Z_-}{Z_+}=\frac{
\sum_{P}
\int\mathcal{D}R(\tau)(-1)^Pe^{-\frac{S_E}{\hbar}}}{
\sum_{P}\int\mathcal{D}R(\tau)e^{-\frac{S_E}{\hbar}}}
\equiv \langle (-1)^P\rangle_+.
\end{equation}
or $J=-k_BT\ln\langle(-1)^P\rangle_+$. Thus the exchange
coupling can be calculated by sampling a symmetric (bosonic) path integral \cite{Ceperley:1995}
and taking the average of $(-1)^P$, which is $+1$ for identity
paths and $-1$ for exchanging paths. 

\begin{figure}[tb]
\includegraphics[width=3.375 in]{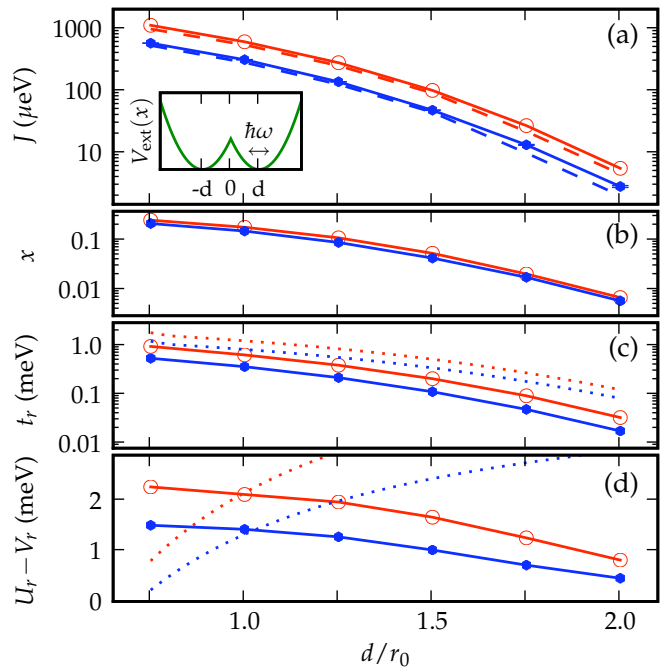}%
\caption{(Color online) PIMC results. (a) Exchange couplings $J$ for $\hbar\omega=4$~meV ($\bullet$) and $\hbar\omega=6$~meV ($\circ$) double QDs
with a piecewise parabolic potential (inset).  Dashed lines are direct diagonalization
results from Ref.~\onlinecite{Pedersen:2007}. (b) The double dot occupation probability $x$. Using $J$ and $x$ we fit (c) $t_r$ and (d) $U_r$ parameters for an effective two site Hubbard-Model. 
Dashed lines in (c) show the bare hopping $t$ for a single electron in the double QD, while the dashed
line (d) is for $V=e^2/\epsilon d$ and $U$ taken from a PIMC calculation on a single QD.
\label{fig:results}}
\end{figure}

We ran PIMC simulations \cite{Ceperley:1995} with our open-source 
{\tt pi} code for the dots studied in Ref.~\onlinecite{Pedersen:2007},
with the results shown in Fig.~\ref{fig:results} (a). While each geometry was sampled
for four hours on eight cores, the algorithm can be ran for just one minute 
to get quick answers with larger error bars. To aid other researchers, we have 
made the simulation available  as a tool on nanoHUB \cite{nanohub:2008}. Coulomb interactions
are included with a crude pair action that correctly handles the cusp condition.
We observed convergence of the path integral results with 6400 discrete
slices, but a higher-quality pair action~\cite{Ceperley:1995} could require fewer slices.
We see near perfect agreement with DD, and speculate that small deviations
are due to the finite  basis in the DD calculation, which can be
a problem as $d$ increases~\cite{Pedersen:2007}.

\begin{figure}[tb]
\includegraphics[width=3.375 in]{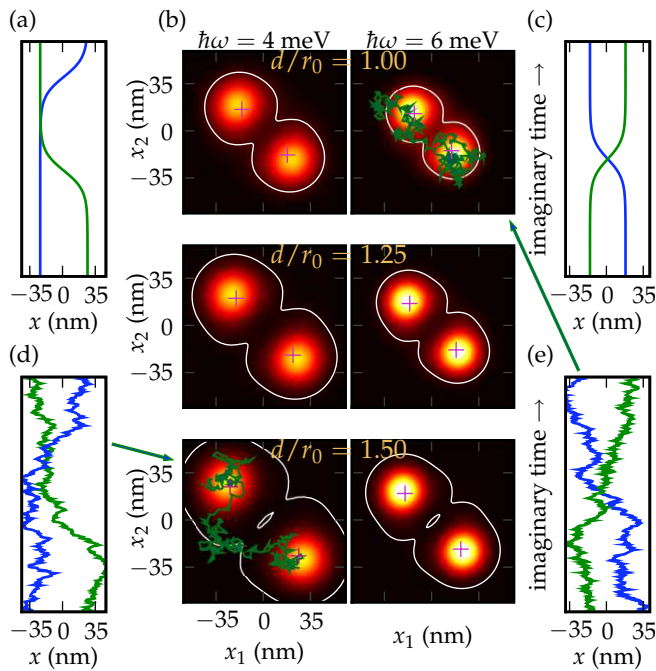}%
\caption{(Color online) Paths and pair densities for a double dot. (a) Simplified instanton with
double occupation of the right dot. (b) Pair densities $\rho(x_1,x_2)$ with the lowest density contour 
line that conects both potential minima (+ markers) at $(\pm d,\mp d)$. (c) Simplified instanton with 
simultaneous exchange.  (d) Actual path showing double occupation, sampled from 
$\hbar\omega=4$~meV, $d=1.5r_0$ QDs. (e) Actual path showing simultaneous exchange,
sampled from $\hbar\omega=6$~meV, $d=2r_0$ QDs. Trajectories (d) and (e) are also plotted in (b).
\label{fig:pairDens}}
\end{figure}

To learn more from our simulation we collect the two-particle density, $\rho(x_1,x_2)$, which is
the probability to find one electron at $x_1$ and the other at $x_2$, integrated over all
values of $y_1$ and $y_2$, shown in Fig.~\ref{fig:pairDens}~(b).
We calculate double occupation, $x_D$, which we define as the probability for 
the electrons to lie on the same side of the $x=0$ plane. 
From $J$ and $x_D$ we are able to deduce renormalized
values for $t_r$ and $U_r-V_r$ in an effective two site model, Fig.~\ref{fig:results} (c) and (d).
Interactions renormalize $t_r$ to smaller values. This is consistent with Hund-Mulliken theory
or a larger renormalized mass. The larger $J$ comes from the dramatic decrease in
$U_r-V_r$ at larger dot separations, as correlation enables more virtual hoping.

There are two minima, $(x_1,x_2)=(\pm d,\mp d)$, in the external potential
$V_{ext}(\mathbf{r}_1)+V_{ext}(\mathbf{r}_2)$, marked `+' in Fig.~\ref{fig:pairDens}(b).
For non-zero $J$, some paths must go between these minima.
In a semiclassical model, the paths fluctuate around the potential minima, with rapid 
crossings called instantons, in which particles exchange between the dots.
An instanton can involve brief double-occupation of a dot, illustrated 
in Fig.~\ref{fig:pairDens}(a), or simultaneous exchange, as in 
Fig.~\ref{fig:pairDens}(c). In Fig.~\ref{fig:pairDens}(d) and (e) we
show actual paths sampled from our simulation that resemble the idealized instantons.
In Fig.~\ref{fig:pairDens}(c), one  instanton can be seen to move from
the $(d,-d)$ minimum, briefly double-occupy the left dot, $(-d,-d)$, then move to the $(-d,d)$ minimum,
while the other instanton moves directly between the two minima. 

\begin{figure}
\includegraphics[width=3.375 in]{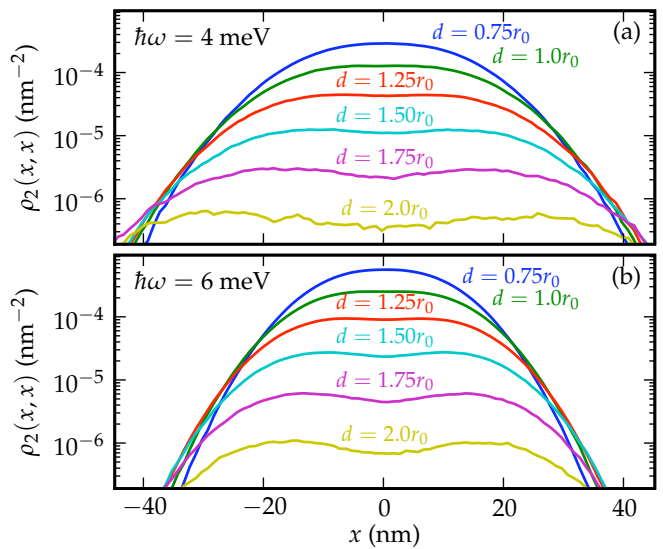}%
\caption{(Color online) Crossing density, $\rho_2(x,x)$, equivalent to the diagonal of the pair densities
in Fig.~\ref{fig:pairDens}(b). \label{fig:pcross}}
\end{figure}

Contours of $\rho(x_1,x_2)$,
Fig.~\ref{fig:pairDens}(b), illustrate a trend with increasing dot separation.
For small $d$ the highest probability
is directly between the minima (simultaneous exchange), but at larger $d$ the highest
probability has two pathways (brief double occupation). 
Fig.~\ref{fig:pcross} show the probability density
for crossing, $\rho(x,x)$. 
Crossing is most likely in the middle ($x=0$) when the dots are close together. When the dots are farther apart, the crossing probability has a double peak near $x=\pm d$, while the probability for crossing in the middle is approximately half the peak value. The double peaks are
slightly larger for the wider $\hbar\omega=4$~meV dot, consistent with more double
occupation, but the difference is not very pronounced. 

\begin{figure}
\includegraphics[width=3.375 in]{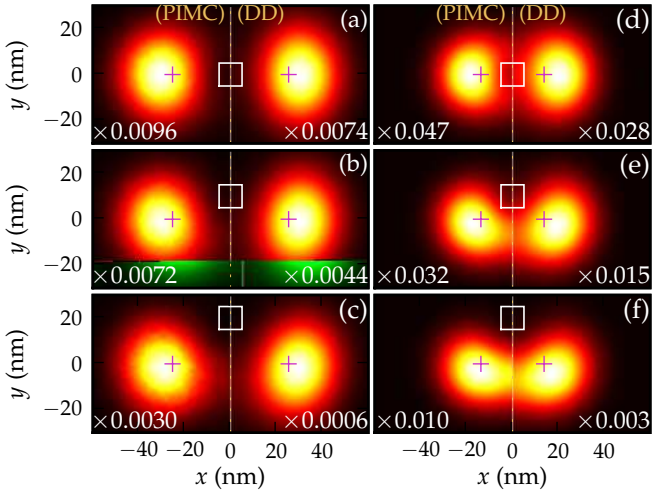}%
\caption{(Color online) Conditional density of one electron when the other electron is in the white box,
showing the correlation hole during an instanton. Panels (a)--(c) are the $\hbar\omega=4$~meV, $d=1.5 r_0$ QDs and (d)-(f) are the $\hbar\omega=6$~meV, $d=1.0 r_0$ QDs.
Numerical factors are the likelyhood of the first electron being in the white box.
PIMC results are shown on the left of each image, with  DD results~\cite{Pedersen:2007}
on the right.
\label{fig:corHole}}
\end{figure}

To underscore the presence of electronic correlation during tunneling, 
we plot the correlation hole form in Fig.~\ref{fig:corHole}, with PIMC results next to
DD results~\cite{Pedersen:2007}. While some quantitative differences are apparent, consistent
with the finite basis size in DD, the overall agreement is quite good. The message is clear:
in the instanton, as one electron moves between the dots, the
other electron moves away,  enhancing the instanton and increasing $J$.

\begin{figure}
\includegraphics[width=3.375 in]{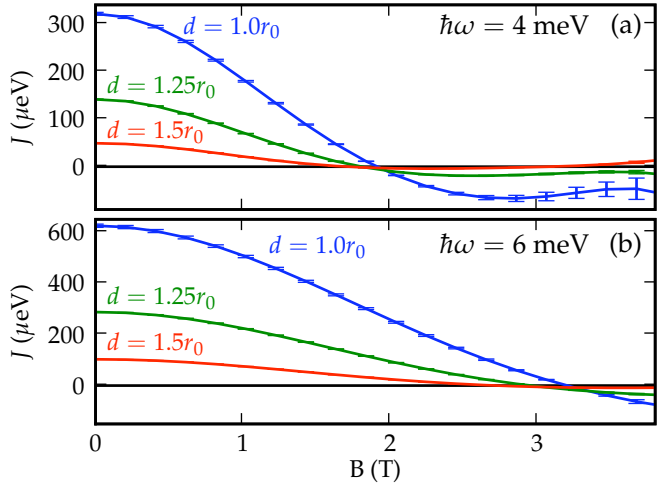}%
\caption{(Color online) Magnetic field dependence included with a Berry's phase for several double QDs.
\label{fig:bfield}}
\end{figure}

For charged particles, magnetic fields can be used to tune the exchange coupling and
even change its sign \cite{Burkard:1999}. In the path 
integral, a magnetic field is easily implemented as a Berry's phase
$q\Phi_B$, where $q$ is the electron charge and $\Phi_B$ is the total magnetic flux enclosed by the worldlines of the two electrons. The exchange splitting is then
$J(B)=-k_BT\ln(\langle e^{iq\Phi_B}(-1)^P\rangle_+/\langle e^{iq\Phi_B}\rangle_+)$.
The quantities are averaged from the bosonic path integral with no field, so data for
different B-fields may be collected simultaneously. For very
large magnetic fields the expectation value in the denominator is small and
Monte Carlo sampling errors are catastrophic. In practice, we find that fields up to 4 T in strength
are practical for the geometries we study, yielding the results in Fig.~\ref{fig:bfield}. 

\begin{figure}
\includegraphics[width=3.375 in]{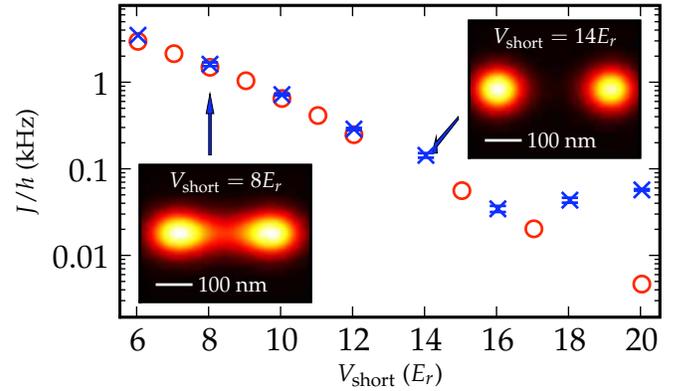}%
\caption{(Color online) Spin-splitting of $^{87}$Rb atoms trapped in a double well:
$\times$, PIMC results at 10 nK, and $\circ$,
experimental data \cite{Trotzky:2008}. Insets show atomic
probability densities.
\label{fig:atoms}}
\end{figure}

As a final example, consider the exchange of two $^{87}$Rb atoms in a double-well optical trap
\cite{Trotzky:2008}. This system resembles the double QD, only with much heavier particles, a much
lower temperature, short range interactions, and a different confining potential. While the experiments
in \cite{Trotzky:2008} have very little correlation, we present results here to show the method is practical
for such systems and could be used for predictions of strongly interacting systems.
The experiment has a double-well potential,
$V(x)=V_{\text{long}}\sin^2(\pi x/\lambda) + V_{\text{short}}\cos^2(2\pi x/\lambda)$, with
$\lambda=765$~nm and $V_{\text{long}}=10E_r$,
where $E_r=h^2/2M_{\text{Rb}}\lambda^2$ \cite{Trotzky:2008}. We model interactions as 
$V(r)=V_0\operatorname{sech}^2 \kappa r$ with $V_0=50.5$~$\mu$K
and $\kappa=0.1$~nm$^{-1}$ to reproduce
the $^{87}$Rb scattering length. Fig~\ref{fig:atoms} shows $J$ as the barrier
$V_{\text{short}}$ is varied, confirming agreement with experiment.

In conclusion, we have demonstrated a PIMC algorithm for computing exchange-splitting in 
double QDs and optical lattices. The exchange splitting arises from instantons in the
path integral, and we have collected statistics for these path crossings, including 
double occupation and the correlation hole. Correlations renormalize $t_r$ and $U_r-V_r$, with
dramatic decrease in $U_r-V-r$ at large separation.  We find that simultaneous crossing occurs 
more often with closely spaced dots, while further separated dots are more likely to have
instantons with double occupations. Finally, we have demonstrated the versatility of the algorithm
with the inclusion of magnetic fields and applications to laser-trapped atoms.

\begin{acknowledgments}
Work supported by NSF Grant No.\ DMR 0239819 and NRI-SWAN and 
made use of facilities provided by the Ira A. Fulton High Performance 
Computing Initiative. JS thanks Erich Mueller  for helpful discussions.
\end{acknowledgments}

\bibliography{DoubleDot}

\begin{thebibliography}{11}
\expandafter\ifx\csname natexlab\endcsname\relax\def\natexlab#1{#1}\fi
\expandafter\ifx\csname bibnamefont\endcsname\relax
  \def\bibnamefont#1{#1}\fi
\expandafter\ifx\csname bibfnamefont\endcsname\relax
  \def\bibfnamefont#1{#1}\fi
\expandafter\ifx\csname citenamefont\endcsname\relax
  \def\citenamefont#1{#1}\fi
\expandafter\ifx\csname url\endcsname\relax
  \def\url#1{\texttt{#1}}\fi
\expandafter\ifx\csname urlprefix\endcsname\relax\def\urlprefix{URL }\fi
\providecommand{\bibinfo}[2]{#2}
\providecommand{\eprint}[2][]{\url{#2}}

\bibitem[{\citenamefont{Mattis}(1981)}]{Mattis:1981}
\bibinfo{author}{\bibfnamefont{D.~C.} \bibnamefont{Mattis}},
  \emph{\bibinfo{title}{The Theory of Magnetism}}, no.~\bibinfo{number}{17} in
  \bibinfo{series}{Springer series in solid-state science}
  (\bibinfo{publisher}{Springer-Verlag}, \bibinfo{address}{Berlin},
  \bibinfo{year}{1981}).

\bibitem[{\citenamefont{Loss and DiVincenzo}(1998)}]{Loss:1998}
\bibinfo{author}{\bibfnamefont{D.}~\bibnamefont{Loss}} \bibnamefont{and}
  \bibinfo{author}{\bibfnamefont{D.~P.} \bibnamefont{DiVincenzo}},
  \bibinfo{journal}{Phys. Rev. A} \textbf{\bibinfo{volume}{57}},
  \bibinfo{pages}{120} (\bibinfo{year}{1998}).

\bibitem[{\citenamefont{Burkard et~al.}(1999)\citenamefont{Burkard, Loss, and
  DiVincenzo}}]{Burkard:1999}
\bibinfo{author}{\bibfnamefont{G.}~\bibnamefont{Burkard}},
  \bibinfo{author}{\bibfnamefont{D.}~\bibnamefont{Loss}}, \bibnamefont{and}
  \bibinfo{author}{\bibfnamefont{D.~P.} \bibnamefont{DiVincenzo}},
  \bibinfo{journal}{Phys. Rev. B} \textbf{\bibinfo{volume}{59}},
  \bibinfo{pages}{2070} (\bibinfo{year}{1999}).

\bibitem[{\citenamefont{Pedersen et~al.}(2007)\citenamefont{Pedersen, Flindt,
  Mortensen, and Jauho}}]{Pedersen:2007}
\bibinfo{author}{\bibfnamefont{J.}~\bibnamefont{Pedersen}},
  \bibinfo{author}{\bibfnamefont{C.}~\bibnamefont{Flindt}},
  \bibinfo{author}{\bibfnamefont{N.~A.} \bibnamefont{Mortensen}},
  \bibnamefont{and} \bibinfo{author}{\bibfnamefont{A.-P.} \bibnamefont{Jauho}},
  \bibinfo{journal}{Phys. Rev. B} \textbf{\bibinfo{volume}{76}},
  \bibinfo{pages}{125323} (\bibinfo{year}{2007}).

\bibitem[{\citenamefont{Ghosal et~al.}(2006)\citenamefont{Ghosal,
  G{\"u\c{c}l\"u}, Umrigar, Ullmo, and Baranger}}]{Ghosal:2006}
\bibinfo{author}{\bibfnamefont{A.}~\bibnamefont{Ghosal}},
  \bibinfo{author}{\bibfnamefont{A.~D.} \bibnamefont{G{\"u\c{c}l\"u}}},
  \bibinfo{author}{\bibfnamefont{C.~J.} \bibnamefont{Umrigar}},
  \bibinfo{author}{\bibfnamefont{D.}~\bibnamefont{Ullmo}}, \bibnamefont{and}
  \bibinfo{author}{\bibfnamefont{H.~U.} \bibnamefont{Baranger}},
  \bibinfo{journal}{Nature Physics} \textbf{\bibinfo{volume}{2}},
  \bibinfo{pages}{336} (\bibinfo{year}{2006}).

\bibitem[{\citenamefont{Berry}(1984)}]{berry:1984}
\bibinfo{author}{\bibfnamefont{M.~V.} \bibnamefont{Berry}},
  \bibinfo{journal}{Proc. R. Soc. A} \textbf{\bibinfo{volume}{392}},
  \bibinfo{pages}{45} (\bibinfo{year}{1984}).

\bibitem[{\citenamefont{Trotzky et~al.}(2008)\citenamefont{Trotzky, Cheinet,
  Folling, Feld, Schnorrberger, Rey, Polkovnikov, Demler, Lukin, and
  Bloch}}]{Trotzky:2008}
\bibinfo{author}{\bibfnamefont{S.}~\bibnamefont{Trotzky}},
  \bibinfo{author}{\bibfnamefont{P.}~\bibnamefont{Cheinet}},
  \bibinfo{author}{\bibfnamefont{S.}~\bibnamefont{Folling}},
  \bibinfo{author}{\bibfnamefont{M.}~\bibnamefont{Feld}},
  \bibinfo{author}{\bibfnamefont{U.}~\bibnamefont{Schnorrberger}},
  \bibinfo{author}{\bibfnamefont{A.~M.} \bibnamefont{Rey}},
  \bibinfo{author}{\bibfnamefont{A.}~\bibnamefont{Polkovnikov}},
  \bibinfo{author}{\bibfnamefont{E.~A.} \bibnamefont{Demler}},
  \bibinfo{author}{\bibfnamefont{M.~D.} \bibnamefont{Lukin}}, \bibnamefont{and}
  \bibinfo{author}{\bibfnamefont{I.}~\bibnamefont{Bloch}},
  \bibinfo{journal}{Science} \textbf{\bibinfo{volume}{319}},
  \bibinfo{pages}{295} (\bibinfo{year}{2008}).

\bibitem[{\citenamefont{Harowitz et~al.}(2005)\citenamefont{Harowitz, Shin, and
  Shumway}}]{Harowitz:2005b}
\bibinfo{author}{\bibfnamefont{M.}~\bibnamefont{Harowitz}},
  \bibinfo{author}{\bibfnamefont{D.}~\bibnamefont{Shin}}, \bibnamefont{and}
  \bibinfo{author}{\bibfnamefont{J.}~\bibnamefont{Shumway}},
  \bibinfo{journal}{J. Low Temp. Phys.} \textbf{\bibinfo{volume}{140}},
  \bibinfo{pages}{211} (\bibinfo{year}{2005}).

\bibitem[{\citenamefont{Feynman}(1972)}]{Feynman:1972}
\bibinfo{author}{\bibfnamefont{R.~P.} \bibnamefont{Feynman}},
  \emph{\bibinfo{title}{Statistical Mechanics}}
  (\bibinfo{publisher}{Addison-Wesley}, \bibinfo{address}{Reading, MA},
  \bibinfo{year}{1972}).

\bibitem[{\citenamefont{Ceperley}(1995)}]{Ceperley:1995}
\bibinfo{author}{\bibfnamefont{D.~M.} \bibnamefont{Ceperley}},
  \bibinfo{journal}{Rev. Mod. Phys.} \textbf{\bibinfo{volume}{67}},
  \bibinfo{pages}{279} (\bibinfo{year}{1995}).

\bibitem[{\citenamefont{Shumway and Gilbert}(2008)}]{nanohub:2008}
\bibinfo{author}{\bibfnamefont{J.}~\bibnamefont{Shumway}} \bibnamefont{and}
  \bibinfo{author}{\bibfnamefont{M.}~\bibnamefont{Gilbert}},
  \emph{\bibinfo{title}{Spin coupled quantum dots}} (\bibinfo{year}{2008}),
  \bibinfo{note}{doi: 10254/nanohub-r4963.1}.

\end{thebibliography}

\end{document}